\documentclass[aps,pra,twocolumn,amsmath,superscriptaddress,longbibliography]{revtex4-2}

\newcommand{\bea}{\begin{eqnarray}}
\newcommand{\eea}{\end{eqnarray}}
\newcommand{\beq}{\begin{equation}}
\newcommand{\eeq}{\end{equation}}

\usepackage{amsmath}
\usepackage[urlcolor=blue,colorlinks=true,citecolor=blue,linkcolor=blue,pdfstartview={FitH},bookmarks=false]{hyperref}
\usepackage{appendix}
\usepackage{graphicx}
\usepackage{longtable}
\usepackage{epsfig}
\usepackage{dcolumn}
\usepackage{bm}
\usepackage{bbm}
\usepackage{amssymb}
\usepackage{multirow}
\usepackage{times,color}
\usepackage{hyperref}
\usepackage{amsmath}
\usepackage{color}
\usepackage{subfigure}
\usepackage{float}
\usepackage{epstopdf}

\begin{document}

 \title{Continuously varying critical exponents  in an  exactly solvable  long-range cluster XY model}

\author {Tian-Cheng Yi}
\affiliation{Department of Physics,
Zhejiang Sci-Tech University, Hangzhou 310018, China}
\author{Chengxiang Ding}
\affiliation{School of Science and Engineering of Mathematics and Physics, Anhui University of Technology,
Maanshan, Anhui 243002, China}

\author{Maoxin Liu}
\affiliation{School of Systems Science  $\&$ Institute of Nonequilibrium Systems, Beijing Normal University, Beijing 100875, China}
\author{Liangsheng Li}
\affiliation{National Key Laboratory of Scattering and Radiation, Beijing 100854, China} 
                   
\author{Wen-Long You}
\email{wlyou@nuaa.edu.cn}
\affiliation{College of Physics, Nanjing University of Aeronautics and Astronautics, Nanjing, 211106, China}
\affiliation{Key Laboratory of Aerospace Information Materials and Physics (NUAA), MIIT, Nanjing 211106, China}

\begin{abstract}
We investigate a generalized antiferromagnetic cluster XY model in a transverse magnetic field, where long-range interactions decay algebraically with distance. This model can be exactly solvable within a free fermion framework. By analyzing the gap, we explicitly derive the critical exponents $\nu$ and $z$, finding that the relationship $\nu z = 1$ still holds. However, the values of $\nu$ and $z$ depend on the decaying exponent $\alpha$, in contrast to those for the quantum long-range antiferromagnetic Ising chain. To optimize scaling behavior, we verify these critical exponents using correlation functions and fidelity susceptibility, achieving excellent data collapse across various system sizes by adjusting fitting parameters. Finally, we compute the entanglement entropy at the critical point to determine the central charge $c$,  and find it also varies with $\alpha$. This study provides insights into the unique effect of long-range cluster interactions on the critical properties of quantum spin systems. 
\end{abstract}

\maketitle
\section{Introduction}
Understanding quantum phase transitions (QPTs) in many-body systems, which describe complex behaviors as external parameters are varied, is fundamental to quantum physics~\cite{sachdev2000}. A widely studied model for exploring QPTs is the transverse Ising model~\cite{pfeuty1970one,pfeuty1971ising}.
Initially developed to describe spin interactions in magnetic materials~\cite{Ising1925Beitrag}, the Ising model finds numerous applications in frontier fields such as quantum computing~\cite{PhysRevA.108.022612, imoto2024universal,PhysRevResearch.3.033108}, quantum information~\cite{dutta2015quantum}, machine learning~\cite{carrasquilla2017machine,Wanglei2016PhysRevB.94.195105,PhysRevA.109.042428}, and social dynamics~\cite{Castellano2009Statistical}.
 While the one-dimensional and two-dimensional versions of the Ising model can be solved exactly~\cite{kramers1941statistics,kramers1941statistics2,statistics1944two}, the three-dimensional case remains unsolved~\cite{PhysRevX.13.021009}. The original Ising model considered only nearest-neighbor interactions.  However, describing many-body systems solely through local interactions is often an approximation that may fail in many real materials, where the long-range interactions  play a crucial role in capturing essential physics~\cite{RevModPhys.95.035002}. 
In particular, long-range interactions can give rise to emergent phases and critical behaviors that differ fundamentally from those in systems dominated by short-range interactions~\cite{PhysRevLett.109.267203,PhysRevLett.111.207202,PhysRevLett.111.260401,PhysRevB.88.045426,juhasz2014random,vodola2015long,PhysRevA.93.053620,
PhysRevB.94.075156,PhysRevA.96.043621,PhysRevA.98.023607,PhysRevB.102.174424,PhysRevB.103.245135,adelhardt2024monte,PhysRevB.110.155120,PhysRevA.110.042408,PhysRevE.110.044121,PhysRevB.110.184502}. 

Unlike typical long-range interactions between two distant sites, cluster models involve interactions spanning multiple sites. The cluster Ising model is another important extension~\cite{PhysRevA.84.022304,PhysRevE.100.042131,PhysRevA.105.053311}, closely related to the cluster state in quantum computation~\cite{schwartz2016deterministic,larsen2019deterministic}.  Cluster state quantum computing~\cite{PhysRevLett.98.190504,raussendorf2007topological,zhou2003quantum,dawson2006noise,nielsen2004optical,nielsen2006cluster,kitaev2003fault} is also known as measurement-based quantum computing, where entangled cluster states are used as the resource for performing quantum computations~\cite{kitaev2003fault}. An increase in the number of entangled qubits has exponentially enhanced quantum computing capabilities~\cite{cao2023generation}.
 
Meanwhile, advancements in quantum simulation techniques have significantly expanded the capabilities  for constructing complex quantum many-body systems and exploring exotic phases of matter~\cite{ebadi2021quantum,semeghini2021probing}.
Rydberg atoms are particularly notable for their exceptional programmability, due to a combination of Van der Waals and dipole-dipole interactions~\cite{PhysRevLett.131.203003,bernien2017probing, scholl2021quantum, bluvstein2021controlling, browaeys2020many, labuhn2016tunable}, and  thus provide an excellent platform for studying such models with cluster interactions. This platform enables precise manipulation of neutral atoms to exhibit long-range and tunable interactions, including long-range dipolar XY interactions~\cite{chen2023continuous, PhysRevB.109.144411}.
The cluster XY model introduces anisotropy as a tunable parameter and has attracted increasing attention~\cite{deger2019geometric}. However, the QPTs and critical phenomena in cluster models remain insufficiently explored. Understanding the universality class of these transitions is essential for uncovering the unique properties of many-body systems, particularly given recent experimental advances in measuring critical exponents~\cite{PhysRevB.99.094203, chepiga2021kibble}.

In this work, we fill this gap by investigating the critical behavior of a generalized antiferromagnetic cluster XY model with long-range interactions that decay algebraically with distance. The model is analytically solvable, which enables the precise extraction of critical exponents. Interestingly, we find that the critical exponents vary continuously with the decay exponent, deviating significantly from behavior observed in typical long-range models. These analytical results are further validated through finite-size scaling analyses of correlation functions and several information-theoretic measures.
 
This paper is organized as follows. In Sec.~\ref{sec2}, we briefly present the long-range cluster XY models and explicitly derive the critical exponents by analyzing the gap. In Sec.~\ref{sec3},
we analyze the critical behaviors of correlation functions. In addition, fidelity susceptibility and entanglement entropy are analyzed.  
The summary and conclusion are given in Sec.~\ref{sec4}.

\section{Model and quantities}
\label{sec2}
 We consider a spin-1/2 long-range cluster XY model in a transverse magnetic field, 
which can be given by 
\begin{eqnarray}
H
&=&\sum_{j=1}^{N} \sum_{m=1}^{M} J_m\left(\frac{1+\gamma}{2}{\sigma}^{x}_{j}{\sigma}^{x}_{j+m}
+\frac{1-\gamma}{2}{\sigma}^{y}_{j}{\sigma}^{y}_{j+m}\right)\nonumber\\
&\times& O_{j+1,m-1}^z -h\sum_{j=1}^{N} \sigma^z_j ,
\label{eq:ham}
\end{eqnarray}
where the operators $\sigma^{x,y,z}_{j}$
are the Pauli matrices for spin at $j$th site, the Jordan-Wigner string  $O_{j+1,m-1}^z$=$\left(\prod_{p=j+1}^{j+m-1}{\sigma}^{z}_{p}\right)$ accumulates the parity of $m-1$ consecutive spins starting from $j+1$th site, $\gamma$ denotes the anisotropy of the XY interactions, and
$h$ characterizes the strength of the transverse magnetic field.
Here the periodic boundary condition (PBC) is assumed, i.e., $\sigma_{N+j}^a= \sigma^a_j$.
The interactions $J_m$ between two spins decay algebraically with
the distance $m$ as
\begin{eqnarray}
\label{eq:decayJ}
J_m = J m^{-\alpha}.
\end{eqnarray}
For convenience, we set $J=1$ and the power-law-decaying exponent $\alpha >0$.
In the cluster XY model~(\ref{eq:ham}), it is
apparent that the ground-state energy of the system diverges for
$\alpha \le 1$ if one does the infinite sums.
The largest distance of long-range interactions is truncated to be $M$.  For the case $M$ = 1, Eq.(\ref{eq:ham}) is reduced to the thoroughly studied anisotropic XY model. For $M$ = 2,  Eq.(\ref{eq:ham}) has  (XZX + YZY) type three-site interactions. The halfway interactions are the farthest interactions correspond to $M$ = $N/2$ (assuming $N$ is even). 
The Hamiltonian in Eq.~(\ref{eq:ham}) exhibits distinct symmetry properties based on the value of $\gamma$. For $\gamma \neq 0$, the system has a discrete $\mathbb{Z}_2$ symmetry. However, when $\gamma = 0$, this symmetry is enhanced to a continuous U(1) symmetry. As a result, the critical properties vary fundamentally between these cases—a distinction that holds not only for the nearest-neighbor Ising case ($M=1$) but also extends to arbitrary $M$ with the inclusion of Jordan-Wigner strings.

\begin{figure}[!htb]
\centering
\includegraphics[width=0.5\textwidth]{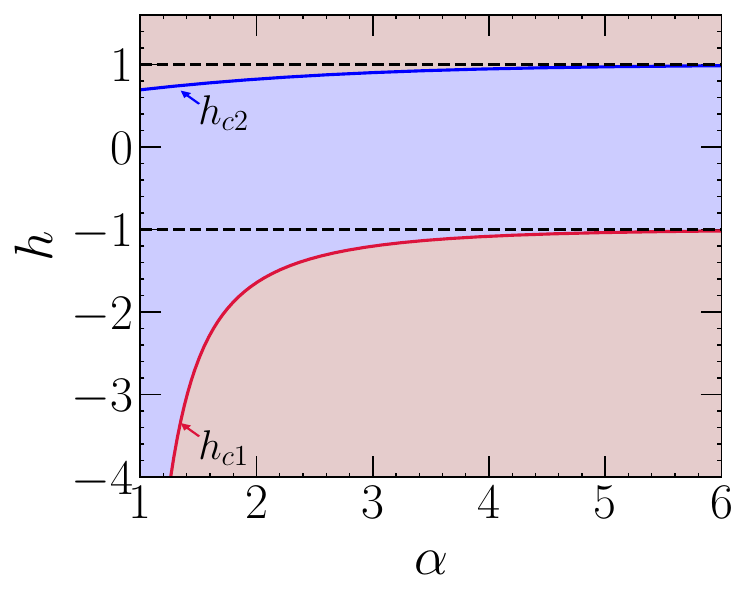}
\caption{
Critical lines for the power-law-decaying exponent $\alpha$ with $M = \infty$, calculated according to  Eq.~(\ref{eq:hc}). The red and blue lines denote the critical fields $h_{c1}$ and $h_{c2}$, respectively. The dashed lines indicate the location of the critical point as $\alpha\rightarrow\infty$.}
\label{fig:hc}
\end{figure}

The Jordan-Wigner transformation maps commuting operators of different spins into anti-commuting fermionic operators as
\begin{eqnarray}
\sigma _{j}^{+}& =&\exp\left[ i\pi\sum_{n=1}^{j-1}c_{n}^{\dagger }c_{n}^{}
\right] c_{j}^{}=O_{1,j-1}^z c_{j}^{},    \\
\sigma _{j}^{-}& =&\exp\left[-i\pi\sum_{n=1}^{j-1}c_{n}^{\dagger }c_{n}^{}
\right] c_{j}^{\dagger}=O_{1,j-1}^z c_j^{\dagger }, \\
\sigma _{j}^{z}& =&1-2c_{j}^{\dagger }c_{j}^{},\label{eq:jw}
\end{eqnarray}
where spin ladder operators are defined as $\sigma^{\pm}_{j}$=$\frac{1}{2}$$(\sigma^x_j\pm i\sigma^y_j)$, and $c_j^+$ ($ c_j$) represents the fermion creation (annihilation) operator at site $j$.
By Jordan-Wigner transformation the Hamiltonian in Eq.(\ref{eq:ham}) can thus be written as a
quadratic form of the creation operator and annihilation
operator of spinless fermions.
\begin{eqnarray}
H_{\rm F}&=&
\sum_{j=1}^{N}
\{\sum_{m=1}^{M}
J_m[(c_{j+m}^{+}c_{j}+c_{j}^{+}c_{j+m}) \nonumber \\
&&+\gamma(c_{j+m}c_{j}+c_{j}^{+}c_{j+m}^{+})]  
-h(1-2c_{j}^{+}c_{j})\}.
\label{eq:ham:jw0}
\end{eqnarray}
Note that the boundary term in Eq. (\ref{eq:ham:jw0}) contains an additional phase factor $c_{N+1}$= $ c_1 (-1)^{(N_p+1)}$, where the total fermion number $N_p$=$\sum_{j=1}^N c_{j}^\dagger c_{j}$.
This subtle boundary effect leads to either  PBC or antiperiodic boundary conditions (APBC) for the spinless 
fermion chain~\cite{lieb1961two}. The boundary contribution becomes negligible in the thermodynamic limit due to the $1/N$ correction. To diagonalize the
 Hamiltonian in Eq.(\ref{eq:ham:jw0}) within the APBC channel, we consider systems with even fermion-number parity.
By performing a Fourier transformation, we obtain the Hamiltonian in momentum space:
\begin{eqnarray}
&H_{\rm F}&=\sum_{k=-\pi}^{\pi}
\left(\begin{array}{cc}c_{k}^{\dag } & c_{-k}\end{array}\right)
\hat{M}_k
\left(\begin{array}{c} c_{k} \\ c_{-k}^{\dag}\end{array}\right),
\end{eqnarray}
with
\begin{equation}
\hat{M}_k =
\left(\begin{array}{cc}
\epsilon_{k} & -i\delta_{k} \\
i\delta_{k} & -\epsilon_{k}
\end{array}\right),
\label{hamfourier}
\end{equation}
and
\begin{eqnarray}
\epsilon_{k}=\sum_{m=1}^{M}J_m\cos{mk}+h, \delta_{k}=\sum_{m=1}^{M}J_m\gamma\sin{mk}, 
\label{delta_k}    
\end{eqnarray}
where $ k = n\pi/N$, 
$n =-(N - 1/2), -(N - 3/2),..., (N - 3/2), (N - 1/2)$.
We diagonalize the Hamiltonian by successive application of the Bogoliubov transformation,
 \begin{eqnarray}
 {\eta}_{k}=\cos(\theta_k/2)c_k-i\sin(\theta_k/2) c_{-k}^{+},
 \end{eqnarray}
 imposing $\theta_{-k}=-\theta_k$. To this end, the Hamiltonian can be reduced to a diagonal form given by
\begin{eqnarray}
H_{\rm F}&=& \sum_{k=-\pi}^{\pi}{\varepsilon}_{k}({\eta}_{k}^{+}{\eta}_{k}-\frac{1}{2}),
\label{Ek}
\end{eqnarray}
where the excitation energy of Bogoliubov
quasiparticles
\begin{eqnarray}
{\varepsilon}_{k}=2\sqrt{\delta_{k}^2+\epsilon_{k}^2}
\end{eqnarray}
 and
the ground state with the form of a BCS state
\begin{eqnarray}
\vert \Psi_g \rangle=\prod_{0\le k\le\pi}[\cos(\theta_k/2)+i\sin(\theta_k/2) c_{k}^{+} c_{-k}^{+}] \vert 0\rangle_c,
\end{eqnarray}
with  
\begin{eqnarray}
\theta_k=-\arctan(\delta_k/\epsilon_k), 
\label{theta_k}
\end{eqnarray}$
\eta_k \vert \Psi_g \rangle$=0, $c_k \vert 0 \rangle_c$=0.
In this way, the ground-state energy can be obtained $E_0\!=\!-\!\frac{1}{2}\!\sum_{k}\!{\varepsilon}_{k}$.
The gap closes when $\delta_{k_c}=0$ and $\epsilon_{k_c}=0$. Two straightforward solutions can be yielded for $k_{c1}=0$ and $k_{c2}=\pi$. The critical values of the magnetic field corresponding to the transitions from the antiferromagnetic phase to the paramagnetic phase are
\begin{eqnarray}
\label{eq:criticalh}
h_{c1}=-\sum_{m=1}^{M} J_m,
h_{c2}=-\sum_{m=1}^{M}{(-1)^m}J_m.
\label{eq:hc}
\end{eqnarray}
Regarding Eq.(\ref{eq:decayJ}) with $M=\infty$, $h_{c1}$=-$\zeta(\alpha)$ and $h_{c2}$=$(1-2^{1-\alpha})\zeta(\alpha)$, where the Riemann zeta function $\zeta(\alpha)$ is a function of a complex variable $\alpha$ that analytically continues the sum of the Dirichlet series $\zeta(\alpha)$=$\sum_{m=1}^\infty$ $ m^{-\alpha}$. For $\alpha \to 1$, $h_{c1}$ goes to negative infinity and $h_{c2}$ approaches $\ln 2 $. Conversely, as $\alpha$  approaches infinity, which corresponds to the antiferromagnetic XY model with nearest-neighbor  interactions, $h_{c1}$ tends toward -1 and $h_{c2}$ approaches 1 (see Fig.~\ref{fig:hc}).

The critical behavior is determined by those low-energy
states near the critical mode. As $h$ approaches, the gap vanishes as 
$\Delta\sim(h-h_c)^{\nu z}$, where $\nu$ and $z$ are the
correlation length and dynamic exponents, respectively.
The gap near criticality $h_{c1}$ is
\begin{equation}
\label{Delta1}
\Delta_{1} \simeq  {\varepsilon}_{0} =2\vert h+\sum J_m \vert=2 \vert h-h_{c1}\vert,
\end{equation}
and the one near criticality $h_{c2}$ is
\begin{equation}
\label{Delta2}
\Delta_{2} \simeq  {\varepsilon}_{\pi} =2\vert h+\sum (-1)^m J_m \vert=2 \vert h-h_{c2}\vert.
\end{equation}
The critical exponents then satisfy $\nu z=1$.
Since the size dependence of the gap, $\Delta \sim N^{-z}$, defines the
dynamic exponent $z$, we expand the gap around the critical line $h_c$
from threshold critical mode $k_c$, i.e., 
$|k-k_c| \ll 1$,
\begin{eqnarray}
\varepsilon_0 &\sim&  -2\left( \sum_{m=1}^{M} m J_m \gamma  \right) |k|,  
\\
\varepsilon_\pi &\sim& -2 \left( \sum_{m=1}^{M} (-1)^{m} m J_m \gamma \right)  |k-\pi|.
\end{eqnarray}
 Take the first critical point $h_{c1}$ as an example,  the critical exponent $z$  can be extracted by the following relationship, 
\begin{eqnarray}
N^{-z}&\sim& 2\vert \sum_{m=1}^{M} m J_m \gamma  \vert \cdot |k|.
\label{eq_extract_z}
\end{eqnarray}

\begin{figure}[!htb]
\centering
\includegraphics[width=0.5\textwidth]{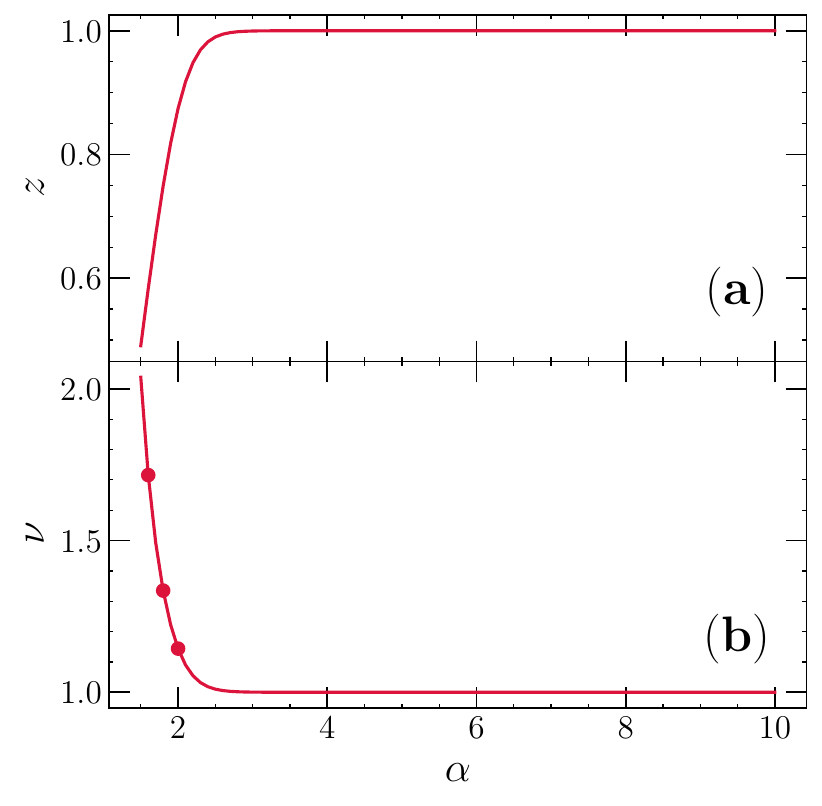}
\caption{
Critical exponents $\nu$ and $z$ at the first critical point $h_{c1}$ as a function of $\alpha$ for the long-range antiferromagnetic cluster XY chain with $\gamma \neq 0$. Three specific values of $\alpha$ ($\alpha=1.6, 1.8, 2.0$) are highlighted with red solid dots, as they will be used in subsequent scaling analysis.
Here we use $N_{\rm min}=1000$ and $N_{\rm max}=10000$ to perform the fitting.
}
\label{fig:nu_z}
\end{figure}

Critical exponent $\nu $ and $z$ for the first critical point $h_{c1}$ with respect to $\alpha$ for 
the long-range cluster XY model are shown in Fig.~\ref{fig:nu_z}. 
To obtain the critical exponent $z$, we selected a range of $N$ values from $N_{\text{min}}$
  to $N_{\text{max}}$
  for fitting.
In this model, the critical exponent is no longer constant and will change with $\alpha$. 
When $\alpha > 3$, $\nu=1$, $z=1$. 
In this case, the model come back to the traditional XY model with only nearest-neighbor interactions.
Notably, the case of $\alpha=2$ represents a special point where the correlation function is characterized by an exponent $\nu=1$, albeit with logarithmic corrections. These corrections may account for the observed numerical deviation of the exponent from the expected value of $\nu=1$~\cite{PhysRevB.100.184306}.
In numerical calculations for extracting the critical exponents, noticeable finite-size effects are observed~\cite{SM}.

\begin{figure}[!htb]
\centering
\includegraphics[width=0.5\textwidth]{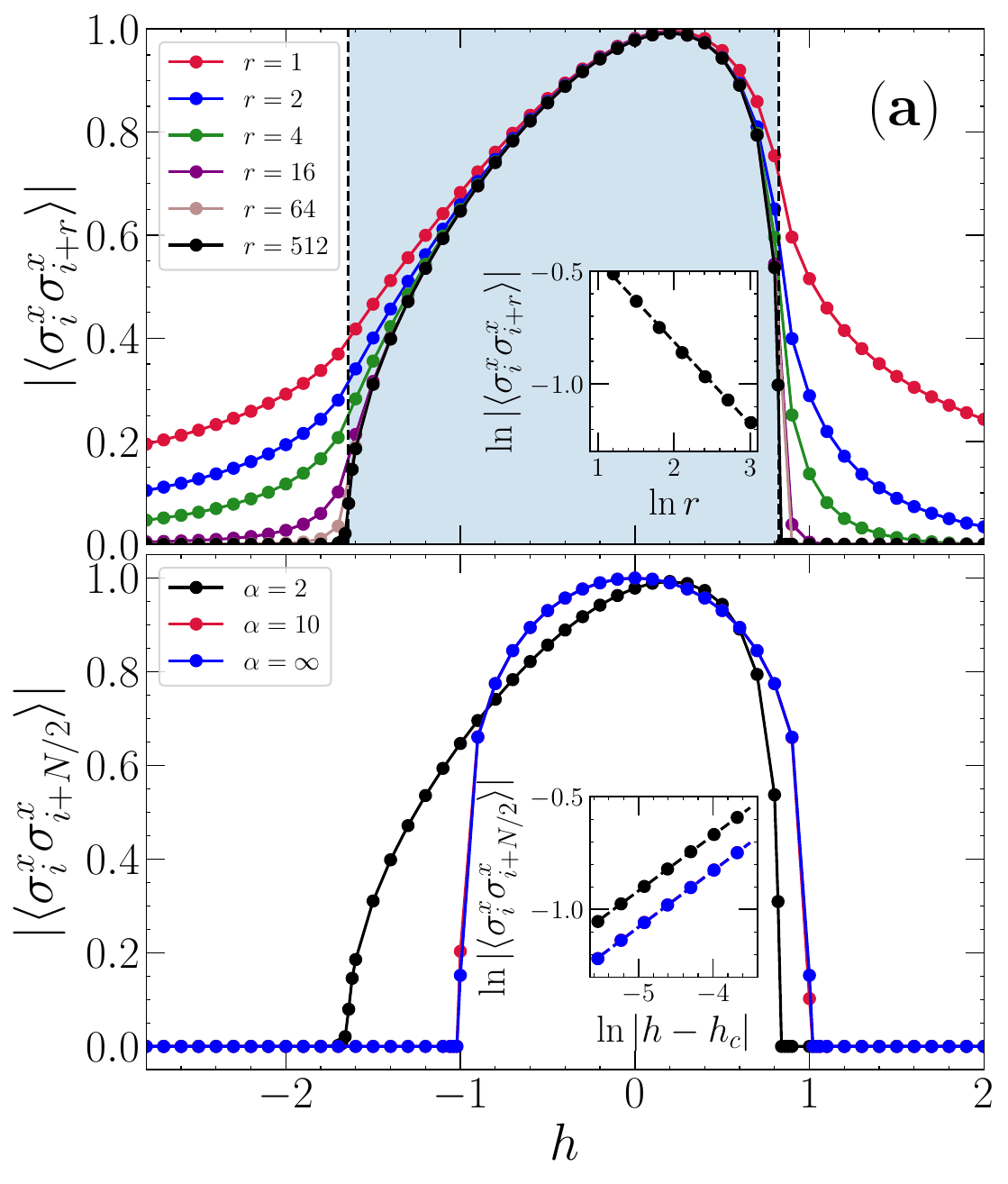}
\caption{
Correlation function $|\langle \sigma_i^x \sigma_{i+r}^x \rangle|$ as a function of $h$ under various conditions:
(a) Correlation for different values of $r$ with $\alpha=2$. The inset displays the power-law decay of the correlation at the critical point $h_{c2}$, with a decay exponent of -0.3647; 
(b) Correlation for different values of $\alpha$ when $r=N/2$. The inset shows the power-law relationship between $\ln |\langle\sigma_i^x\sigma_{i+N/2}^x\rangle|$ and $\ln |h-h_c|$, with exponents of 0.2461, 0.2498, and 0.2498 for $\alpha=2$, 10, and $\infty$, respectively. 
Here, we set $\gamma=1$ and $N=1024$.
}
\label{fig:LongRangeOrder}
\end{figure}

\begin{figure}[!htb]
\centering
\includegraphics[width=0.5\textwidth]{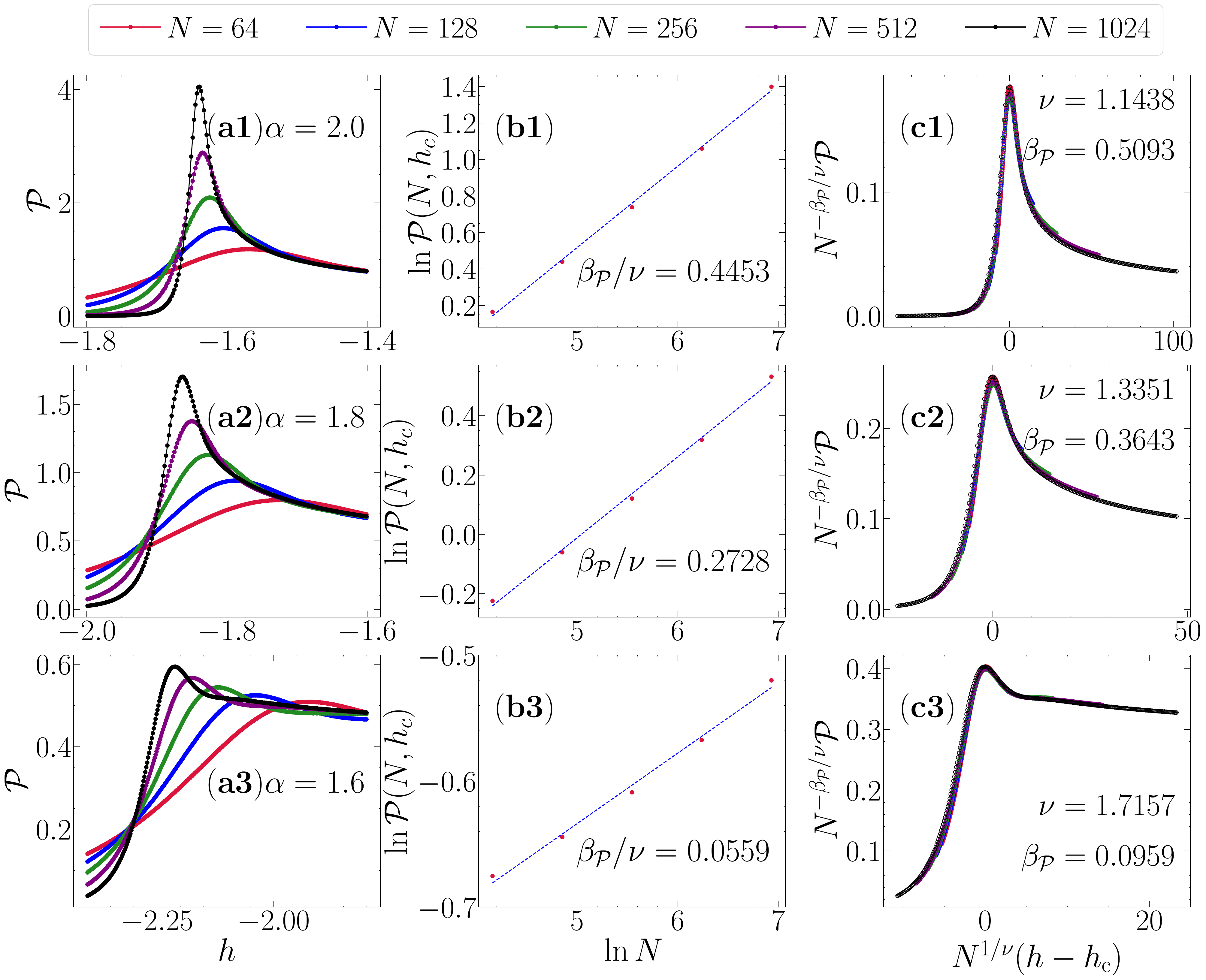}
\caption{
Critical behavior of the correlation function for $\alpha=2.0$, $1.8$, and $1.6$ with $\gamma=1.0$: 
(a) First derivative of the correlation function ${\cal P}$ with respect to $h$;
(b) Logarithmic plot of ${\cal P}(N,h_c)$ versus system size $N$;
(c) Scaled derivative of the correlation function, $N^{-\beta_{\cal P}/\nu} {\cal P}$, plotted as a function of the scaled variable $N^{1/\nu}(h-h_c)$. 
All curves for different lattice sizes collapse onto a single curve when the correlation length critical exponent is set to $\nu = 1.1438$, $1.3351$, and $1.7157$ for $\alpha=2.0$, $1.8$, and $1.6$, respectively.
}
\label{fig:scaling_xx}
\end{figure}
 \section{Results}
\label{sec3}

To determine whether the ground state of the model has long-range order, we calculate 
the two-point correlations $\langle{\sigma}_{i}^{x}{\sigma}_{i+r}^{x}\rangle$.
From the Jordan-Wigner transformation, we have
\begin{eqnarray}
\langle{\sigma}_{i}^{x}{\sigma}_{i+r}^{x}\rangle
&=&\langle
B_{i}A_{i+1}
B_{i+1}A_{i+2}
\cdots
B_{i+r-1}A_{i+r}
\rangle,\nonumber\\
\end{eqnarray}
with
$A_i=c_i^+ + c_i$
and
$B_i=c_i^+ - c_i$.
When $\varepsilon_k$ is always positive,
this correlation function can be calculated by the Wick theorem~\cite{lieb1961two,tian2018quantum,PhysRevB.100.024423,PhysRevE.102.032127,liu2021quantum,PhysRevA.105.063306}, namely 
\begin{eqnarray}
&&\langle{\sigma}_{i}^{x}{\sigma}_{i+r}^{x}\rangle
\nonumber
\\&=&\left|
\begin{array}{cccc}
\langle B_{i}A_{i+1}\rangle & \langle B_{i}A_{i+2}\rangle & \cdots & \langle B_{i}A_{i+r}\rangle \\
\langle B_{i+1}A_{i+1}\rangle & \langle B_{i+1}A_{i+2}\rangle & \cdots & \langle B_{i+2}A_{i+r}\rangle \\
\vdots & \vdots & \ddots & \vdots \\
\langle B_{j-1}A_{i+1}\rangle & \langle B_{j-1}A_{i+2}\rangle & \cdots & \langle B_{j-1}A_{i+r}\rangle
\end{array}
\right|,  
\nonumber\\
\label{eq:Gxx}
\end{eqnarray}
where
\begin{eqnarray}
\langle B_{i}A_{i+r}\rangle
=-\frac{1}{2\pi}\int_{-\pi}^{\pi}\frac{\cos(rk)\epsilon_{k}- \sin(rk)\delta_{k}
}{\sqrt{\epsilon_{k}^2+\delta_{k}^2}} dk.
\end{eqnarray}%
 
As shown in Fig.~\ref{fig:LongRangeOrder}(a), the correlation function $|\langle \sigma_i^x \sigma_{i+r}^x \rangle|$   is examined for varying distances $r$ as a function of the external magnetic field $h$. In the interval $h_{c1} < h < h_{c2}$, $|\langle \sigma_i^x \sigma_{i+r}^x \rangle|$   tends towards a nonzero constant with increasing $r$, indicating the presence of long-range order characterizing the antiferromagnetic phase. Conversely, for $h < h_{c1}$  or $h > h_{c2}$, the correlation function also saturates but reflects a paramagnetic phase lacking long-range order. Notably, at the critical point $h_{c2}$, the inset reveals a power-law decay of the correlation function as a function of distance $r$, described by $|\langle \sigma_i^x \sigma_{i+r}^x \rangle| \sim r^{k_0}$
    with $k_0 = -0.3647$.

As depicted in Fig.~\ref{fig:LongRangeOrder}(b), by increasing $\alpha$, the region of the antiferromagnetic phase has shrunk.
  It is observed that increasing $\alpha$ results in a shrinkage of the antiferromagnetic phase region, with no discernible differences between $\alpha = 10$ and $\alpha = \infty$. 
 The inset illustrates a power-law relationship between $\ln |\langle \sigma_i^x \sigma_{i+N/2}^x \rangle|$ 
  and $\ln |h - h_c|$, yielding coefficients of 0.2461, 0.2498, and 0.2498 for $\alpha = 2, 10, \infty$, respectively. This comprehensive analysis highlights the dependence of the system's phase behavior on the tuning parameter $\alpha$ and delineates the phase boundaries in the context of the cluster XY model.

The scaling behavior of 
the first derivative of the farthest two-point correlation function in $x$-direction 
$\partial\langle \sigma_i^x \sigma_{i+N/2}^x \rangle/\partial h$
is shown in Fig.~\ref{fig:scaling_xx}. 
To simplify the description, we first define ${\cal P}\equiv\partial\langle \sigma_i^x \sigma_{i+N/2}^x \rangle/\partial h$. 
%
The first derivative of the correlation function ${\cal P}$ for a finite system of size $N$ in the neighborhood of a quantum critical point shall obey the universal scaling form~\cite{Fisher1972Scaling}
\begin{eqnarray}
 {\cal P}(N,h)=N^{\beta_{\cal P}/\nu}  f_{\cal P}(|h-h_c| N^{1/\nu}), 
\label{ansatz_xx}
\end{eqnarray}
where $\nu$ is the correlation-length critical exponent,  $\beta_{\cal P}$ is a fitting parameter chosen to improve data collapse in the scaling analysis, and $f_{\cal P}$ is a universal scaling function.   
Numerical results for ${\cal P}$ as a function of $h$
at $\gamma=1$ for different system sizes are presented in Fig.~\ref{fig:scaling_xx}(a).
As $h$ increases, $P$ exhibits a peak, suggesting the occurrence of a phase transition at this point.  
With increasing system sizes $N$, the peaks of ${\cal P}$ become
more pronounced and the maximal value of ${\cal P}$ is
expected that
\begin{eqnarray}
    {\cal P}(N, h_c) = N^{\beta_{\cal P}/\nu} f_{\cal P}(0).
\end{eqnarray}
When $\alpha=\infty$, $\beta_{\cal P}=2$, $\nu=1$~\cite{PhysRevA.98.023607, PhysRevA.82.012308, PhysRevA.105.013315}. 
As shown in Fig.~\ref{fig:scaling_xx}(b), $\beta_{\cal P}/\nu=0.4453, 0.2728, 0.0559$ for $\alpha=2.0, 1.8, 1.6$ respectively.

\begin{table}[htp!]
\caption{
Critical exponents $\{z, \nu\}$
and order parameter exponents $\{\beta_{\cal P}, {\beta_{F}}\}$ through relations Eqs.~(\ref{ansatz_xx}) and (\ref{ansatz_fs}).}
\label{Fittingparameter}
\begin{ruledtabular}
\begin{tabular}{ c c c c c c}
	$\alpha$	& $\gamma$& $\nu$&$z$&$\beta_{\cal P}$&${\beta_{F}}$ \\
\hline
2.0 &1.0 &1.1438&0.8743&0.5093&0.7056\\
2.0 &0.8 &1.1438&0.8743&0.4852&0.7198\\
2.0 &0.6 &1.1438&0.8743&0.4491&0.7328\\
1.8 &1.0 &1.3351&0.7490&0.3643&0.5402\\
{1.8} &
{0.8} &
{1.3351} &
{0.7490} &
{0.3239} &
{0.5515}\\
{1.8} &
{0.6} &
{1.3351} &
{0.7490} &
{0.2611} &
{0.5557}
\\
1.6 &1.0 &1.7157&0.5829&0.0959&0.2449
\end{tabular}
\end{ruledtabular}
\end{table}

\begin{figure}[!htb]
\centering
 \includegraphics[width=0.5\textwidth]{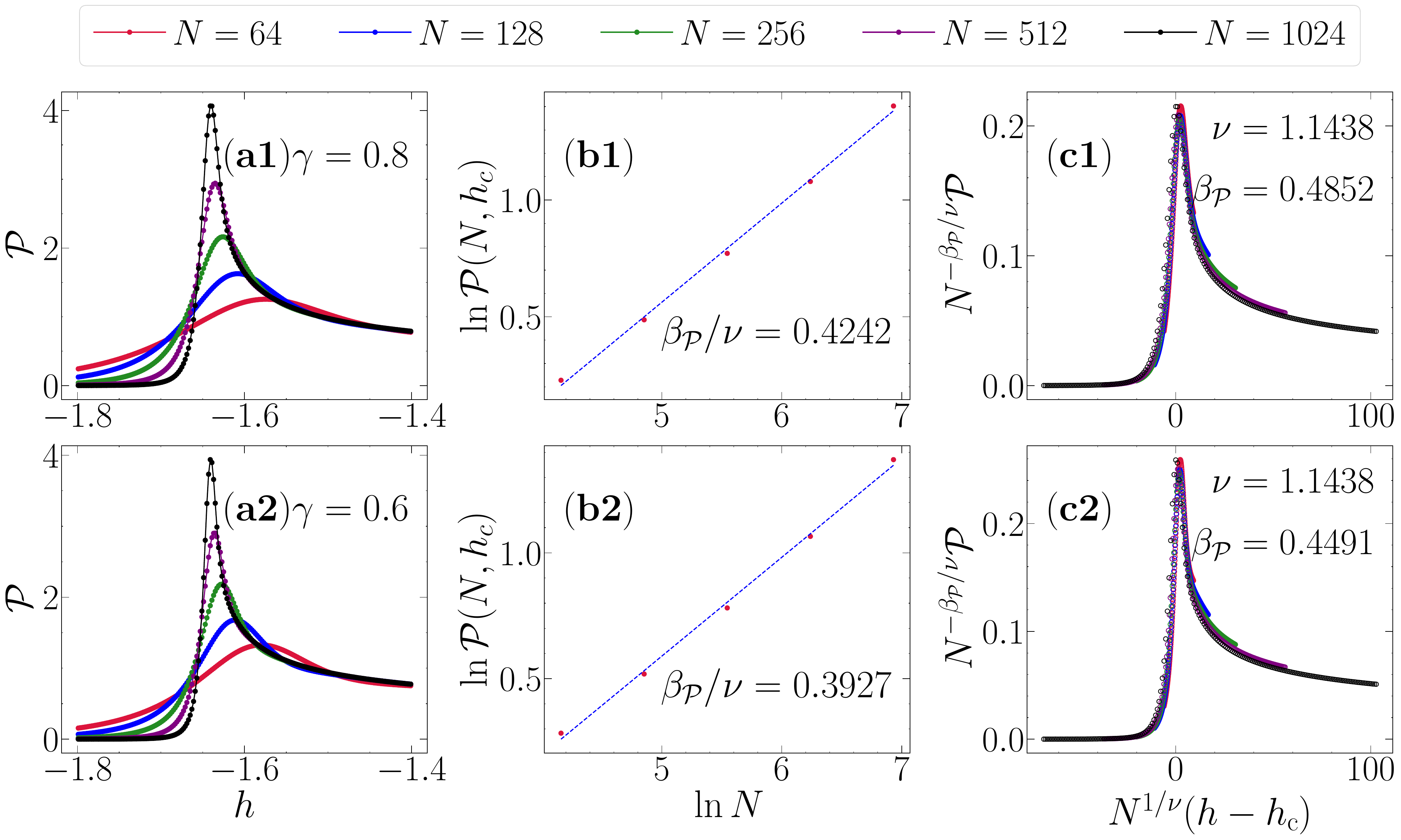}
\caption{
Critical behavior of the correlation function ${\cal P}$ for $\gamma=0.8$ and $0.6$ with $\alpha=2.0$: 
(a) First derivative of the correlation function ${\cal P}$ with respect to $h$;
(b) Logarithmic plot of ${\cal P}(N, h_c)$ versus system size $N$;
(c) Scaled function $N^{-\beta_{\cal P}/\nu}{\cal P}$ as a function of the scaled variable $N^{1/\nu}(h-h_c)$. 
}
\label{fig:scaling_xx_gamma}
\end{figure}

As shown in Fig.~\ref{fig:scaling_xx}(c), 
using the critical exponent  from Eq.~(\ref{eq_extract_z}), we can get a good collapse of scaled function $N^{-\beta_{\cal P}/\nu}{\cal P}$ as a function of scaled variable $N^{1/\nu}(h-h_c)$ for a given $\beta_{\cal P}$, which is extracted from the fitting in Fig.~\ref{fig:scaling_xx}(b).
For $\gamma=0.8$ and $0.6$, similar results are shown in Fig.~\ref{fig:scaling_xx_gamma}. The corresponding critical exponents ${z, \nu}$ and fitting parameters ${\beta_{\cal P}, \beta_F}$ are summarized in Table~\ref{Fittingparameter}.  Different values of $\gamma$ share the same critical exponents $\nu$ and $z$, only with different values of $\beta_{\cal P}$.

\begin{figure}[!htb]
\centering
\includegraphics[width=0.5\textwidth]{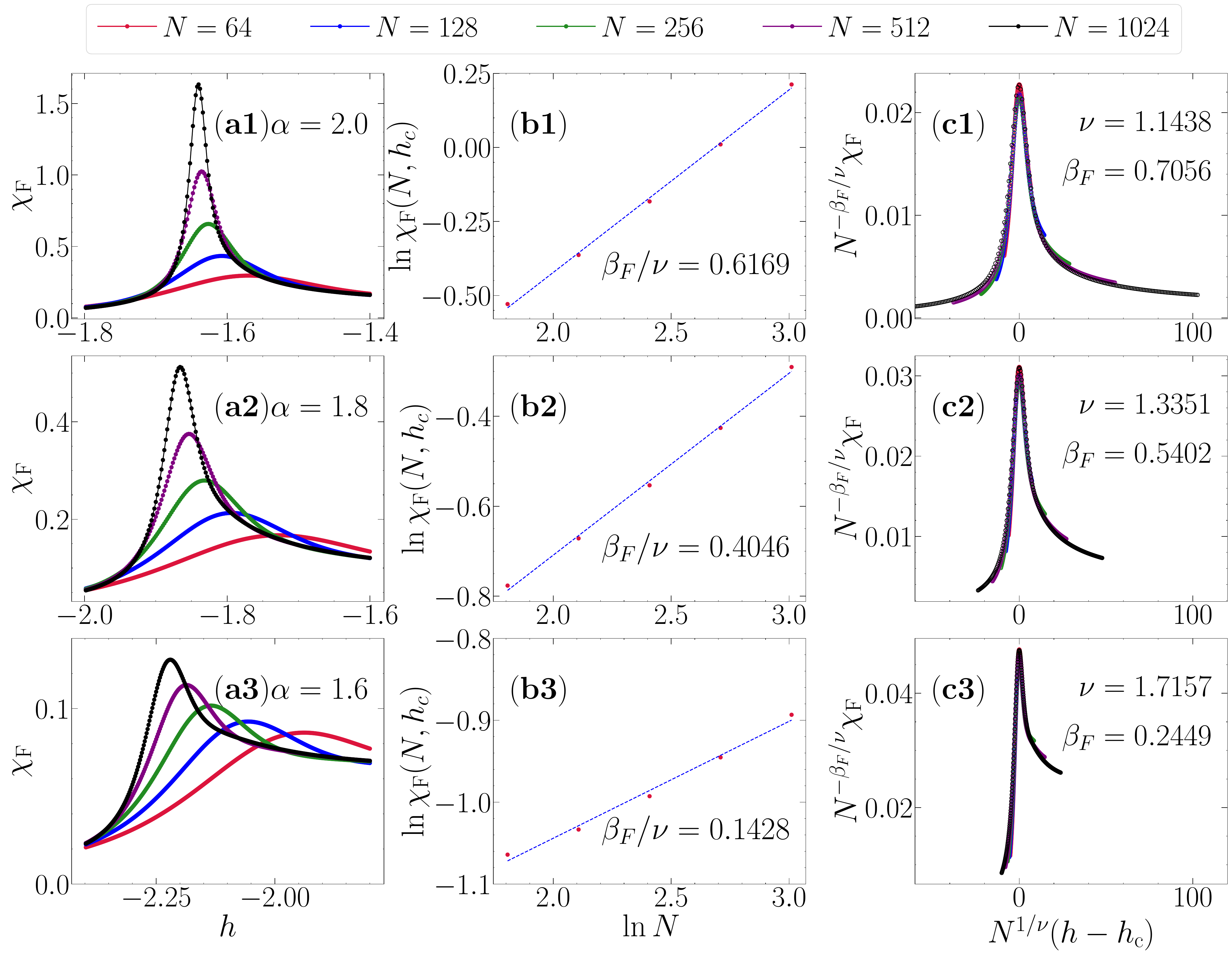}
\caption{ Critical behavior of the fidelity susceptibility $\chi_{\rm F}$ for $\alpha=2.0$, $1.8$, and $1.6$ with $\gamma=1$. 
(a) Fidelity susceptibility $\chi_{\rm F}$ as a function of $h$; 
(b) Logarithmic plot of $\chi_{\rm F}(N, h_c)$ versus system size $N$; 
(c) Scaled fidelity susceptibility $N^{-\beta_F/\nu} \chi_{\rm F}$ as a function of the scaled variable $N^{1/\nu}(h-h_c)$. 
Data for different lattice sizes collapse onto a single curve when choosing the correlation length critical exponents $\nu=1.1438$, $1.3351$, and $1.7157$, respectively.
}

\label{fig:scaling_FS}
\end{figure}

To be more unbiased, we also show the results of the fidelity susceptibility.
The fidelity susceptibility is a general probe of the QPTs.
By definition, quantum fidelity of a many-body Hamiltonian
$\hat{H}(\lambda)=H_0 + \lambda H_I$ is \cite{PhysRevA.75.032109}
\begin{eqnarray}
F(\lambda_0,\lambda_1)=
\vert\langle\Psi_0(\lambda_0)\vert\Psi_0(\lambda_1)\rangle\vert,
\end{eqnarray}
where $\vert \Psi_0 \rangle$ is the ground state, $\lambda_0$ and
$\lambda_1$ specify two points in the parameter space of driving
parameter $\lambda$. In this respect, fidelity susceptibility is
defined as leading order of the Taylor expansion of the overlap
function $F(\lambda,\lambda+\delta \lambda)$, given by
\cite{PhysRevE.76.022101, PhysRevLett.99.095701}
\begin{eqnarray}
\chi_F= \lim_{\delta \lambda \to 0}
\frac{-2 \ln F(\lambda,\lambda+\delta \lambda)}{(\delta \lambda)^2}.
\label{chiform}
\end{eqnarray}
The fidelity susceptiblity $\chi_{\rm F}$ obey the similar universal scaling form 
\begin{eqnarray}
 \chi_{\rm F}(N,h)=  N^{{\beta_{F}}/\nu} f_{F}(|h-h_c| N^{1/\nu}), 
\label{ansatz_fs}
\end{eqnarray}
 where ${\beta_{F}}$ is a fitting parameter used to achieve optimal scaling collapse for the fidelity susceptibility, and $f_{\rm F}$ is a universal scaling function.
Figure~\ref{fig:scaling_FS} shows the scaling behavior for the fidelity susceptibility $\chi_{\rm F}$, yielding the same critical exponents $\nu$ and $z$. 

\begin{figure}[!htb]
\centering
\includegraphics[width=0.5\textwidth]{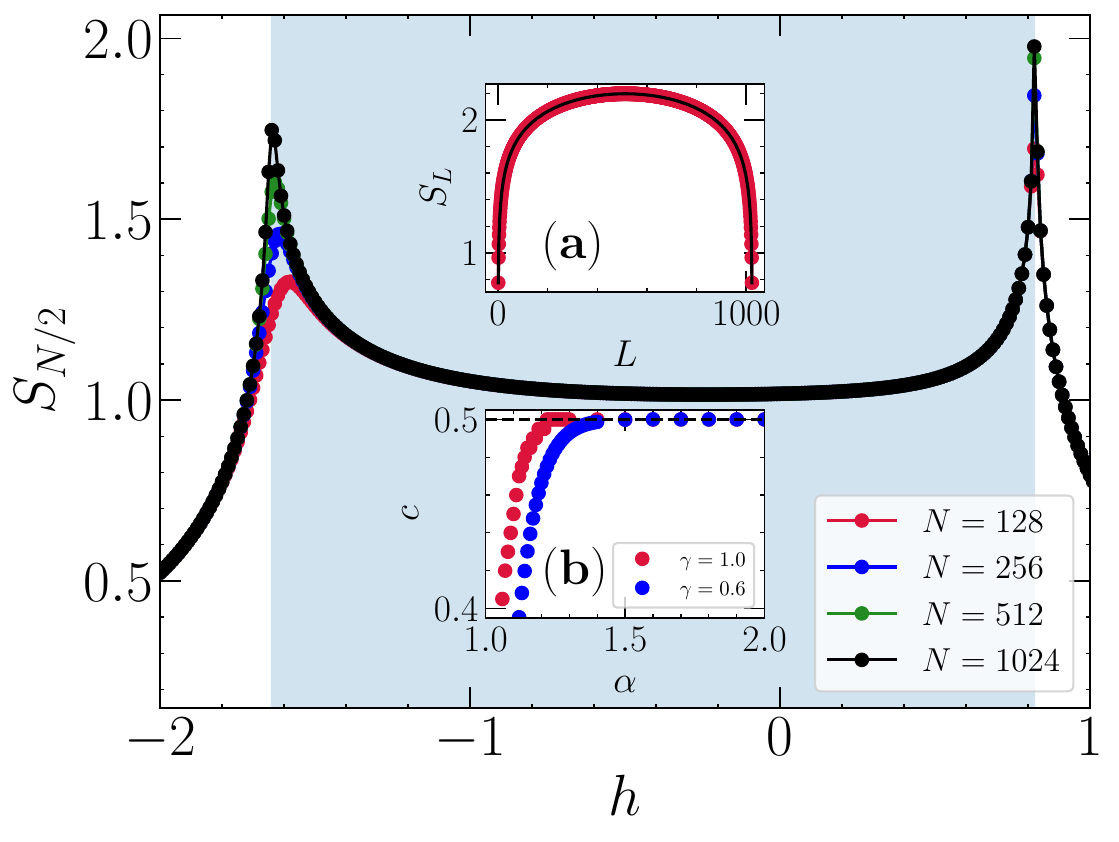}
\caption{
The half-chain entanglement entropy $S_{N/2}$ as a function of $h$ for different system sizes $N$ with $\alpha=2$ and $\gamma=1$. 
Inset (a) displays $S_L$ versus $L$ at the critical point $h_{c2}$, following Eq.~(\ref{s2}). 
Inset (b) shows the central charge $c$ as a function of $\alpha$ at the right critical points for $\gamma=1$ and $\gamma=0.6$.
}
\label{entangle}
\end{figure}

Moreover, we also study the critical behavior of the entanglement entropy.  
Quantum phase transitions arise when quantum fluctuations alter the ground-state properties of a many-body system. At the critical point, entanglement entropy reveals non-local correlations across all distances, which makes it a valuable tool for identifying these transitions. 
The entanglement entropy is defined as follow 
\begin{eqnarray}
S_{L}&=&-\sum_k \left[ \xi_k \ln \xi_k+(1-\xi_k) \ln (1-\xi_k)\right],
\label{s1}
\end{eqnarray}
where $\xi_k$ is the $k$-th eigenvalue of the correlation matrix $\hat{G}_{ij}={\rm Tr}(\rho_A c_i^\dag c_j)$.
Near the critical point, the block entanglement obtains a logarithmic correction with~\cite{calabrese2009entanglement,PhysRevE.105.034128}
 \begin{equation} S_{L} \sim \frac{c}{3}\text{ln}(\frac{N}{\pi}{\rm{sin}}(\frac{\pi L}{N})) + S^{\prime}, 
\label{s2} 
 \end{equation} where $c$ is the central charge, varying for different universality classes, and $S^{\prime}$
 is a non-universal constant, which is shown in the upper inset. 
In the cluster XY model, central charge $c$ is non-universal and varies with $\alpha$.
The block entanglement entropy $S_{L}$ with $N=$ $128$, $256$, $512$, $1024$  along the line which crosses two different phases when $\alpha=2$ is shown in Fig.~\ref{entangle}. Here the half-chain entanglement entropy $S_{N/2}$
can also be treated as a detector of the quantum critical points in the cluster XY model. 
It is observed that the entanglement entropy reaches its maximum at two critical points, which approximately occur at $h \approx -1.64$ and $h \approx 0.82$. 
These peaks in $S_{N/2}$  indicate QPTs of the system.
The inset in Fig.~\ref{entangle}(a) displays the entanglement entropy $S_L$ as a function of the subsystem size $L$ for the system size $N = 1024$. The entropy $S_L$ exhibits a characteristic dome-like shape, peaking in the middle of the system and decreasing towards the edges, consistent with Eq.~(\ref{s2}) which is the prediction for a critical system near a conformal field theory (CFT) describing the phase transition. 
The inset in Fig.~\ref{entangle}(b) shows the dependence of the central charge $c$ on the decaying exponent $\alpha$ for $\gamma=1.0$ and $\gamma=0.6$, with a fixed system size of $N = 1024$. The central charge differs between values of $\gamma$, and as $\alpha$ increases, $c$ gradually converges to approximately $0.5$, suggesting a transition into a critical phase that can be described by conformal field theory (CFT)~\cite{PhysRevA.72.022318}.

\section{Summary}
\label{sec4}

In conclusion, we have analyzed the critical properties of a generalized antiferromagnetic cluster XY model in a transverse magnetic field with algebraically decaying long-range interactions. By using exact solutions within the framework of free fermions, we explicitly determined the critical exponents $\nu$ and $z$ from the energy gap, confirming the relationship $\nu z = 1$. However, unlike the quantum long-range antiferromagnetic Ising chain, where critical exponents remain fixed, we observed that $\nu$ and $z$ vary with the decay exponent $\alpha$. 
To further validate these critical exponents, we examined correlation functions and fidelity susceptibility. By carefully adjusting scaling parameters, we achieved an excellent data collapse onto a single curve for the rescaled field-derivative of the farthest two-point correlation function across different system sizes, demonstrating consistency in our findings. Fidelity susceptibility exhibits a comparable scaling behavior, further supporting these results. 
Additionally, entanglement entropy analysis at the critical point revealed that the central charge $c$ depends on $\alpha$, converging to 0.5 as $\alpha$ increases—a behavior that parallels $z$ and highlights the distinctive effects of cluster interactions on quantum correlations and critical phenomena. Our findings deepen the understanding of non-local interactions in many-body systems and offer insights for designing quantum simulators and computational models leveraging cluster-based interactions to explore complex quantum behaviors.

 \begin{acknowledgments}
This work is supported by the National Natural Science Foundation of
China (NSFC) under Grant No. 12174194, 12404285, Postgraduate Research \&
Practice Innovation Program of Jiangsu Province, under Grant No. KYCX23\_0347, Opening Fund of the Key Laboratory of Aerospace Information Materials and Physics (Nanjing University of Aeronautics and Astronautics), MIIT, Top-notch Academic Programs Project of Jiangsu Higher Education Institutions (TAPP), and stable supports for basic institute research under Grant No. 190101. 
 T.-C.Y. acknowledges support from the Science Foundation of
 Zhejiang Sci-Tech University Grant No. 23062182-Y.
\end{acknowledgments}


\appendix
\bibliography{refs}

\bibliographystyle{apsrev4-2}

\clearpage
\widetext
\setcounter{equation}{0}
\setcounter{figure}{0}
\setcounter{table}{0}
\setcounter{section}{0}
\setcounter{tocdepth}{0}

\numberwithin{equation}{section} 
\begin{center}
{\bf \large Supplemental Material for “Continuously varying critical exponents  in an  exactly solvable  long-range cluster XY model” }
\end{center}

\section{Finite-size effect on the fitted critical exponents}
In this supplementary material, we discuss the impact of finite-size effects on the extraction of the critical exponents.
As we mentioned before, the critical exponent $z$  can be extracted by the following relationship, 
\begin{eqnarray}
N^{-z}&\sim& 2\vert \sum_{m=1}^{M} m J_m \gamma  \vert \cdot |k|, 
\label{eq_extract_z_app}
\end{eqnarray}
where $M=N/2$.
Taking the logarithm of both sides of the above relationship (\ref{eq_extract_z_app}), we obtain: 
\begin{eqnarray}
-z \log N&\sim& 
\log \left( \sum_{m=1}^{M=N/2} \frac{1}{m^{\alpha-1}}\right)+\log  (2 \gamma|k|).
\end{eqnarray}
Here, we can extract the critical exponent $z$ through linear regression.
We choose $N_{\rm min}=N/10$, and $N_{\rm max}=N$ to do the fitting. 

\begin{figure}[!htb]
\centering
\includegraphics[width=0.5\textwidth]{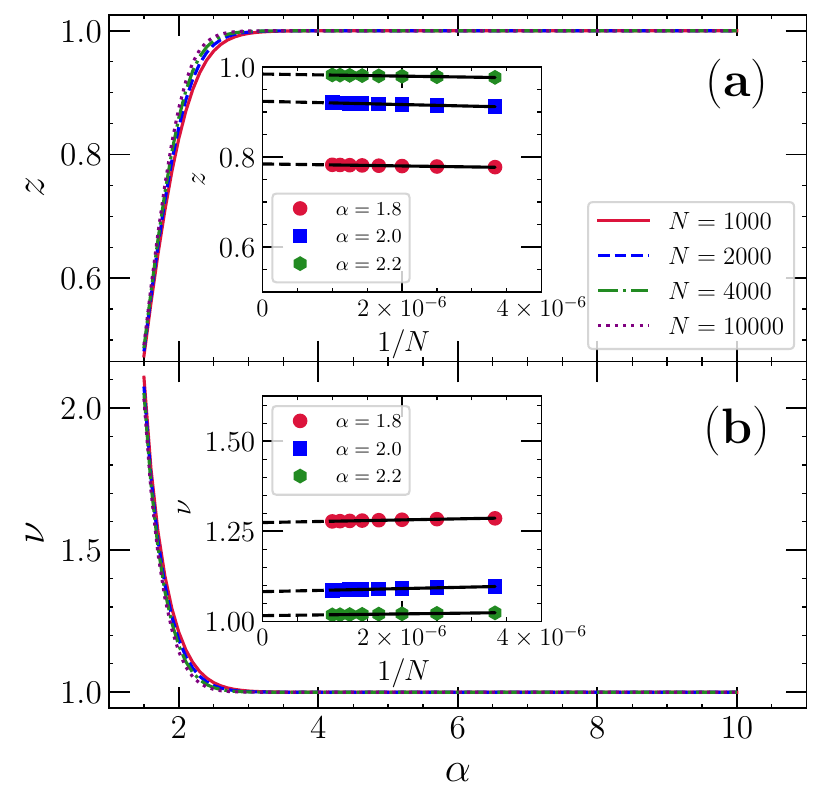}
\caption{
The impact of finite-size effects on the extraction of the critical exponent $\nu$. For the fitting, we select  $N_{\rm min}=N/10$, and $N_{\rm max}=N$ to do the fitting. 
}
\label{sizeeffect}
\end{figure}

The corresponding result is shown in Fig.~\ref{sizeeffect}.
As the system size $N$ increases, the critical exponent $\nu$ will converge to 1.2743, 1.0825, 1.0159 and $z$ will converge to 0.7848, 0.9238, 0.9843   when $\alpha=1.8, 2.0, 2.2$, respectively.

\section{Asymptotic Analysis of the Fitted Critical Exponents}
This scaling behavior can be understood analytically by examining the asymptotic form of  (\ref{eq_extract_z_app}). Specifically, we define $f(N)$ as
\begin{eqnarray}
f(N) \propto \left| \frac{1}{N} \sum_{m=1}^{N/2} \frac{m}{m^\alpha} \right| = \left| \frac{1}{N} H_{N/2, \alpha - 1} \right|,
\end{eqnarray}
where $H_{n,r} = \sum_{j=1}^{n} j^{-r}$  represents the generalized harmonic number for an exponent $r$. To understand the scaling behavior of $H_{n,r}$, we apply the Euler-Maclaurin formula for $n \gg 1$ and $r \neq 1$, which gives the asymptotic expansion, which gives the asymptotic expansion:
 \begin{eqnarray}
H_{n,r} = \zeta(r) + \frac{n^{1-r}}{1-r} + \frac{n^{-r}}{2} - \frac{rn^{r-1}}{12} + O(n^{-r-2}).
\end{eqnarray}
where $\zeta(r)$ is the Riemann zeta function. Substituting this expansion into $H_{N/2, \alpha-1}$, we obtain:
\begin{eqnarray}
H_{N/2,\alpha-1} \approx \zeta(\alpha - 1) + \frac{(N/2)^{2-\alpha}}{2-\alpha} + \frac{(N/2)^{1-\alpha}}{2} - \frac{(\alpha-1)(N/2)^{-\alpha}}{12},
\end{eqnarray}

For large $N$, $H_{N/2, \alpha - 1}$ is dominated by either $\zeta(\alpha - 1)$ or the power-law term $(N/2)^{2-\alpha}$, depending on the value of $\alpha$. Substituting $H_{N/2, \alpha - 1}$ back into $f(N)$, we find:
\begin{eqnarray} f(N) \propto \left| \frac{1}{N} \left[ \zeta(\alpha - 1) + \frac{(N/2)^{2-\alpha}}{2-\alpha} \right] \right|. \end{eqnarray}

For $\alpha > 2$, the term $(N/2)^{2-\alpha}$ becomes asymptotically negligible compared to $\zeta(\alpha - 1)$, so that: \begin{eqnarray} f(N) \propto N^{-1}, \end{eqnarray} yielding $z = 1$. For $\alpha < 2$, $\zeta(\alpha - 1)$ is asymptotically negligible compared to $(N/2)^{2-\alpha}$, leading to: \begin{eqnarray} f(N) \propto N^{1-\alpha}, \end{eqnarray} which implies $z = \alpha - 1$.

These results are consistent with the observed numerical trends.  For $ \alpha \ge  2 $, the first term dominates, while for $ \alpha < 2 $, the second term becomes more significant.  The crossover at $ \alpha = 2 $ marks a transition between the dominance of $ \zeta(\alpha - 1) $ and $ (N/2)^{2-\alpha} $. In this case, logarithmic corrections may arise and require careful consideration in the numerical analysis.

In summary, the asymptotic analysis leads to the following results for the value of $ z $ in the thermodynamic limit:
\begin{eqnarray}
z = 
\begin{cases} 
1 & \text{if } \alpha \geq 2, \\ 
\alpha - 1 & \text{if } \alpha < 2. 
\end{cases} 
\label{eq:z}
\end{eqnarray}

\end{document}